\documentstyle[12pt]{article}
\newcommand{\mypacs}[1]{\noindent PACS:{#1} \\[.1in]}
\newenvironment{mytitle}{\begin{center} \Large \bf}{\end{center}}
\newenvironment{myauthor}{\begin{center} \normalsize }{\end{center}}
\newenvironment{myaddress}{\begin{center} \it }{\end{center}}
\renewcommand{\thesection}{\Roman{section}.}

\newcommand{\acknowledgements}{\renewcommand{\thesection}{}
\section{\hspace{-1em}Acknowledgements}}
\newcounter{myeqnum}
\newenvironment{myeqnarray}{\setcounter{myeqnum}{\value{equation}}
   \stepcounter{myeqnum} \setcounter{equation}{0}
   
   \begin{eqnarray}}{\end{eqnarray} \setcounter{equation}{\value{myeqnum}}
   
   \par \vspace{-1\baselineskip} \noindent \hspace{-0.35em}}
\newcounter{myfigure}
\newenvironment{figurecaptions}{\newpage\section*{\hspace{-1em}Figure captions}
\begin{list}{\bf Fig. \arabic{myfigure}.}{\usecounter{myfigure}
\labelwidth1.27cm \leftmargin1.67cm \labelsep0.4cm \rightmargin1cm
\parsep0.5ex plus0.2ex minus0.1ex \itemsep0ex plus0.2ex}}{\end{list}}
\begin{document}
\mypacs{73.20.Dx, 73.20.Mf, 73.40.Lq, 73.61.Ey}
\begin{mytitle}
Band-gap engineering in two-dimensional periodic photonic crystals
\end{mytitle}
\begin{myauthor}
Manvir S. Kushwaha
\end{myauthor}
\begin{myaddress}
Institute of Physics, University of Puebla, P.O. Box J-48, Puebla 72570,
Mexico.\\
\end{myaddress}
\begin{abstract}
A theoretical investigation is made of the dispersion characteristics of
plasmons in a two-dimensional periodic system of semiconductor (dielectric)
cylinders embedded in a dielectric (semiconductor) background. We consider
both square and hexagonal arrangements and calculate extensive band structures
for plasmons using a plane-wave method within the framework of a local theory.
It is found that such a system of semiconductor-dielectric composite can give
rise to huge full band gaps (with a gap to midgap ratio $\approx 2$) within
which plasmon propagation is forbidden. The most interesting aspect of this
investigation is the huge lowest gap occurring below a threshold frequency and
extending up to zero. The maximum magnitude of this gap is defined by the
plasmon frequency of the inclusions or the background as the case may be. In
general we find that greater the dielectric (and plasmon frequency) mismatch,
the larger this lowest band-gap. Whether or not some higher energy gaps appear,
the lowest gap is always seen to exist over the whole range of filling fraction
in both geometries. Just like photonic and phononic band-gap crystals,
semiconducting band-gap crystals should have important consequences for designing
useful semiconductor devices in solid state plasmas.
\end{abstract}
\newpage
\section{INTRODUCTION}
Ever since the discovery of quantum Hall effect, the investigation of
electronic transport in the low-dimensional systems has witnessed many
important and fascinating results, both theoretical and
experimental$^{1}$. The ever increasing interest in these systems is
attributed partly to advancement in the modern semiconductor technology
(e.g., molecular beam eptaxy and electronic beam lithography) that makes
it possible to fabricate the nanostructures with controllable chemical
composition and geometric structures$^{2}$--- quantum-wells, -wires, 
-dots, -rings, -crossbars, -balls are a few to name. The keystone behind
the intense research interest in such systems world-wide is esentially
the fact that their dimension(s) are (or have to be) comparable to the
de Broglie wavelength, so as to observe the quantum-size effects.

Many novel discoveries in physics have originated from the study of wave 
propagation in periodic structures. Examples include x-ray and electron
diffraction by crystals, electronic band structure, Bloch oscillator, and
holography. Analogies between subfields of physics have also opened 
amazingly new fruitful avenues in research. An exciting example is the 
recent discovery$^{3}$ of periodic dielectric structures that exhibit
what is called "photonic band gap", by analogy with the electronic
band gap in semiconductor crystals. Photons within the frequency range 
of the gaps are completely forbidden, so that atoms in such materials are
denied to spontaneously absorb and re-emit light in this region; this has
obvious implications for highly efficient lasers. Since the interesting
phenomena emerging from photonic band gap  structures, such as the
inhibition/prohibition of spontaneous emission, the formation of
localized states of light, and the photon-atom bound states, are all
the consequences of the existence of the gap(s), a major part of research
efforts has concentrated on the search for such photonic crystals. Within
almost a decade the field has prospered much more than Eli Yablonovitch and
Sajeev John could have imagined$^4$.

It did not take long before photonic crystals aroused interest in the
analogous studies in other systems as well; we refer to the ones
allowing, for example, elastic/acoustic waves$^5$, and spin waves$^6$.
The composite sytems exhibiting complete or pseudogaps, within which 
the respective waves are forbidden, are useful both from the practical
and fundamental point of view. From the practical point of view, such
systems can be used to design the filters that prohibit the propagation
of respective waves at certain frequencies while allowing free
propagation at others. From the fundamental point of view, the
propagation of such waves in a slightly disordered composite system can
lead to the novel phenomenon concerned with the Anderson localization.
It is noteworthy that the authentic seeds of classical wave localization
were disseminated in a classic paper by Sajeev John in 1984$^7$. A
systematic account of the theoretical and experimental work on electronic-,
phononic-, photonic-, and vibrational-band gap crystals is given in an
extensive review article by the author$^4$.

In this work we are concerned with a two-dimensional (2D) periodic system
of semiconducting (dielectric) cylinders  embedded in a dielectric
(semiconductor) background. In fact, it could be considered a bulk-mass
of semiconductor with periodically punched infinitely long thin holes.
The cross-sections of these cylinders (or punched holes) are considered
to be much less than the de Broglie wavelength, so as to neglect the
quantum-size effects. We believe that such geometric structures should
be easily realizable through the aforesaid nanostructure growth
techniques. Our principal interest is the prospect of achieving a complete
plasmonic band gap (PBG); this is defined to be a stop band within which
the plasmon propagation is prohibited irrespective of the direction of
propagation and of the polarization of the plasma wave. The existence of
PBG in such composites should have important consequences for designing
filters for the plasma waves and for the fundamental physics interest
associated with the Anderson localization thereof. In fact, the
possibility of experimental realization of the Anderson localization of
plasma waves in random and incommensurate layered systems was predicted
almost a decade ago$^8$.

It is noteworthy that our model theory is based on the full vectorial
Maxwell's equations. We also consider an applied magnetostatic filed
oriented parallel to the axis of the cylinders (which is taken to be the
z-axis). In order to calculate the band structure of such a binary
composite, we employ a plane-wave method and Fourier-series-expansion of
the position-dependent material parameters. It should thus be pointed
out that we do not make use of any messy electromagnetic boundary
conditions usually employed for inhomogeneous systems. The semiconductor
medium in the inclusions and/or background is characterized by the
frequency-dependent (local) dielectric function. The first hand effort
of using local theory for photonic crystals has been made in the recent
past$^9$. We neglect the damping effects and hence ignore the absorption.
In the state-of-the-art high quality semiconductor systems available,
this is thought to be quite a reasonable approximation. Thus our purpose
is to calculate the band structure of a neat and clean 2D periodic
binary semiconductor-dielectric composites.

A number of investigators have calculated such dispersion characteristics
of 1D periodic systems (superlattices, for example) both with and without
an applied magnetic field$^{10-12}$. However, no one, to our knowledge,
has ever embarked on the possibility of observing complete band gaps in
the systems considered. The authors were, most of the time, interested in
reporting the physical conditions leading to the existence of the bulk
bands and the surface plasmons or magnetoplasmons in the infinite and
semi-infinite superlattice systems. This seems to be true irrespective of
the systems (compositional or doping (n-i-p-i) superlattices) considered,
methodology (classical or quantal) employed, and the orientation of the
magnetic field (Voigt-, perpendicular-, or Faraday-geometry) pondered.
The same remark is valid for the work on the 1D lateral (quantum wire)
superlattices$^{13-14}$.

The rest of the paper is designed as follows. In Sec. II we derive the
secular equations required to compute the plasmon band structures of the
2D binary semiconductor composites. The extensive numerical computational
results are discussed in Sec. III. This includes results both on square
and hexagonal arrangements. A brief summary in Sec. IV concludes our
finding in this paper.

\section {METHODOLOGICAL DETAILS}

We consider a 2D periodic system of semiconducting cylinders embedded in
a different semiconducting background. For the sake of generality, we
consider both cylindrical inclusions and the background materials to be
semiconductors to be characterized by the frequency-dependent (local)
dielectric functions. An external magnetostatic field ($\vec{B_o}$) is
assumed to be oriented along the axis ($\hat{z}$) of the cylinders. We are
interested in the nonmagnetic materials, which implies that
$\vec{B}\equiv \vec{H}$ in the Maxwell equations. In the situation at
hand ($\vec{B_o}\parallel \hat{z}$),the dielectric tensor
$\tilde{\epsilon}$ is simplified by the symmetry requirements such that:
$\epsilon_{xx}=\epsilon_{yy}$, $\epsilon_{yx}=-\epsilon_{xy}$, and
$\epsilon_{zx}=\epsilon_{xz}=\epsilon_{zy}=\epsilon_{yz}=0$.

We start with full vectorial Maxwell's curl field equations. After
eliminating the magnetic field variable $\vec{B}$, we obtain the
following wave field equation for the macroscopic electric field
vector($\vec{E}$):

\begin{equation}
\vec{\nabla}\times(\vec{\nabla}\times\vec{E}) - q^2_0
\tilde{\epsilon}\cdot \vec{E}=0
\end{equation}
where $\tilde{\epsilon}$ is a $3\times3$ matrix and $q_0=\omega/c$ is the
vacuum vector; $c$ being the speed of light in vacuum and $\omega$ is the
angular frequency of the wave. Eq.(1) can be written in the matrix form
as follows:

\begin{equation}
\left [
\begin{array}{lrr}
q^2_0 \epsilon_{xx} + \partial^2_y + \partial^2_z &
q^2_0-\partial_x\partial_y &  - \partial_x\partial_z\\
-q^2_0 \epsilon_{xy} - \partial_y\partial_x & q^2_0 \epsilon_{xx} +
\partial^2_x + \partial^2_z &   - \partial_y\partial_z\\
- \partial_z\partial_x & - \partial_z \partial_y  & q^2_0 \epsilon_{zz}
- \partial^2_x + \partial^2_y\\
\end{array}
\right ]
\left [
\begin{array}{c}
E_x\\
E_y\\
E_z
\end{array}
\right ]
=
\left [
\begin{array}{c}
0\\
0\\
0
\end{array}
\right ]
\end{equation}
Here $\partial_z \equiv \partial/\partial_z$, etc. Eq. (2) can be solved
for three different situations: (i) $k_z=0$, and $\vec{k_\parallel}\ne 0$;
(ii) $k_z\ne 0$ and $\vec{k_\parallel}=0$; and (iii) $k_z\ne 0$ and
$\vec{k_\parallel}\ne 0$. We are here interested in the first case ---
with $k_z = 0$ but $\vec{k_\parallel}\ne 0$. Here $\vec{k_{\parallel}}$ is
the 2D Bloch vector and $k_z$ is the $\hat{z}$-component of the propagation
vector.

As such, eq. (2) splits in two independent sets of equations (with $k_z=0$):
\begin{equation}
\nabla^2 E_z + q^2_0 \epsilon_{zz} E_z = 0
\end{equation}
and
\begin{myeqnarray}
(q^2_0 \epsilon_{xx} + \partial^2_y)E_x + (q^2_0 \epsilon_{xy} -
\partial_x\partial_y)E_y = 0\\
(-q^2_0 \epsilon_{xy} - \partial_x\partial_y)E_x +
(q^2_0\epsilon_{xx} + \partial^2_x)E_y = 0
\end{myeqnarray}
where $\epsilon_{ij}$ are all functions of position. Evidently, eq. (3)
represents the propagation of the plasma waves which remain independent of
the magnetic field; whereas the set of eqs. (4) refers to the
magnetoplasma waves. We will therefore derive two independent secular
equations corresponding to the two situations. It is, however,
noteworthy that both equations (3) and (4) are valid for arbitrary
inhomogeneity.

\subsection{Magnetic-filed-independent waves}

Here we are concerned with the plasma waves governed by eq. (3). We focus
our attention on an inhomogeneous medium which exhibits spatial periodicity.
That means that the position-dependent material parameters
$\epsilon_{zz}(\vec{r})$ may be expanded in the Fourier series. We write
\begin{equation}
\epsilon_{zz}(\vec{r})=\sum_ {\vec{G}}\epsilon_{zz}(\vec{G})  e^{i\vec{G}\cdot
\vec{r}}
\end{equation}
where $\vec{G}$ is the reciprocal lattice vector that has the corresponding
dimensionality. The z-component of the field vector must satisfy the Bloch
theorem:
\begin{equation}
E_z(\vec{r},t)=e^{i(\vec{K}\cdot\vec{r}-\omega t)}\sum_{\vec{G}}E_z(\vec{G})
e^{i\vec{G}\cdot \vec{r}}
\end{equation}
Equations (5) and (6) must be substituted in eq. (3). A straightforward algebra,
then multiplying the whole equation by $exp(-i\vec{G}\cdot \vec{r})$, and
integrating over the unit cell yields the following result:
\begin{equation}
\sum_{\vec{G}'}q^2_0 \epsilon_{zz}(\vec{G}-\vec{G}') E_z(\vec{G}') -
\left |\vec{K}+\vec{G} \right |^2 E_z(\vec{G})=0 \cdot
\end{equation}
For binary composites every unit cell is composed of only two materials ---
inclusions (i) and background (b) characterized by $\epsilon^i_{zz}$ and
$\epsilon^b_{zz}$. In addition, we must specify the filling fraction $f$
and $(1-f)$, respectively, for materials $i$ and $b$. We consider
$\epsilon_{zz}$ to be given by
\begin{equation}
\epsilon_{zz}(\vec{G})=\frac {1}{A_c} \int d^2r \epsilon_{zz}(\vec{r})
e^{-i\vec{G}\cdot \vec{r}}
\end{equation}
where $A_c$ is the area of the unit cell and the integration is over the
unit cell. A simple conceptual aspect of the system leads us to define
\begin{equation}
\epsilon_{zz}(\vec{G})=\left \{
\begin{array}{lll}
\epsilon^i_{zz} f + \epsilon^b_{zz} (1-f)\equiv \overline {\epsilon_{zz}}\ \ ,
\ \mbox{for} \ \vec{G}=0 \\
(\epsilon^i_{zz} - \epsilon^b_{zz}) F(\vec{G})\equiv \Delta \epsilon_{zz}
F(\vec{G})\ \ , \ \mbox{for} \ \vec{G} \ne 0
\end{array}
\right .
\end{equation}
where the structure factor $F(\vec{G})$ is defined as
\begin{equation}
F(\vec{G}) = \frac{1}{A_c} \int_i d^2r e^{-i\vec{G}\cdot \vec{r}} =
2fJ_1(Gr_0)/(Gr_0)
\end{equation}
where $J_1(x)$ is the Bessel function of the first kind of order one, the
integration is limited to a cylinder of radius $r_0$, and the filling
fraction
\begin{equation}
f= \pi r^2_0/A_c = \left \{
\begin{array}{lll}
\pi (r_0/a)^2 \ \ , \ \mbox{for square lattice} \\
\frac {2\pi}{\surd{3}}(r_0/a)^2 \ \ , \ \mbox{for hexagonal lattice}
\end{array}
\right .
\end{equation}
where $a$ is the period (or lattice constant) of the system. With the aid
of eq. (9), one can cast eq. (7) in the form (see Appendix):
\begin{equation}
S \sum_{\vec{G}'\ne \vec{G}} F(\vec{G} - \vec{G}')E_z(\vec{G}') +
\left [ Q + c^2\left |\vec{K} + \vec{G}\right | \right ] E_z(\vec{G}) =
\omega^2 \left [ R\sum_{\vec{G}'\ne \vec{G}} F(\vec{G} - \vec{G}')
E_z(\vec{G}') + P E_z(\vec{G}) \right ]
\end{equation}
where
\begin{myeqnarray}
P = \epsilon^i_L f + \epsilon^b_L (1 - f)\\
Q = \epsilon^i_L \omega^2_{pi} f + \epsilon^b_{pb} \omega^2_{pb}(1-f)\\
R = \epsilon^i_L - \epsilon^b_L\\
S = \epsilon^i_L \omega^2_{pi} - \epsilon^b_L \omega^2_{pb}
\end{myeqnarray}
where $\epsilon^j_L$ and $\omega_{pj}$ are, respectively, the background
dielectric constant and the screened plasma frequency of the medium
concerned; $j\equiv i, b$.

Equation (12) can be cast in the form of a standard eigenvalue problem:
\begin{equation}
\sum_{\vec{g}'} {A}_{\vec{g},\vec{g}'} E_z(\vec{g}')=\Omega^2 E_z(\vec{g}),
\end{equation}
where the dimensionless $\Omega=\omega(a/2\pi c)$, $\vec{k}= \vec{K}(a/2\pi)$,
and $\vec{g}=\vec{G}(a/2\pi)$. The matrix $\stackrel{\leftrightarrow}{A}$
is defined as follows:
\begin{equation}
\stackrel{\leftrightarrow}{A} = \stackrel{\leftrightarrow}{B}^{-1}
\stackrel{\leftrightarrow}{C}
\end{equation}
and the matrix elements of $\stackrel{\leftrightarrow}{B}$ and
$\stackrel{\leftrightarrow}{C}$ are given by
\begin{myeqnarray}
B_{\vec{g},\vec{g}'} = P\delta_{\vec{g},\vec{g}'} + R F(\vec{g}-\vec{g}')
(1-\delta_{\vec{g},\vec{g}'}) \\
C_{\vec{g},\vec{g}'} = \left [ Q' + \left |\vec{k}+\vec{g} \right |^2 \right ]
\delta_{\vec{g},\vec{g}'} + S' F(\vec{g}-\vec{g}')(1-\delta_{\vec{g},\vec{g}'})
\end{myeqnarray}
where
\begin{myeqnarray}
Q' = Q (a/2\pi c)^2 = \epsilon^i_L \Omega^2_{pi} f
+ \epsilon^b_L \Omega^2_{pb}(1 - f)\\
S' = S (a/2\pi c)^2 = \epsilon^i_L \Omega^2_{pi}
- \epsilon^b_L \Omega^2_{pb}
\end{myeqnarray}
where $\Omega_{pj}=\omega_{pj}(a/2\pi c)$ is the normalized screened plasma
frequency. Equation (14) is, in fact, performed at the computational level for
both the geometries considered in the present work. The existence of matrix
$\stackrel{\leftrightarrow}{A}$ implies that its elements must be computed for
every value of $\vec{k}$, rather than for every value of $\vec{k}$ and $\Omega$
as implied by eq. (12). A drastic saving in computational time is achieved by
transforming eq. (12) into eq. (14). The penality one has to pay for this
benefit is that one is required to invert matrix
$\stackrel{\leftrightarrow}{B}$ and perform the multiplication
$\stackrel{\leftrightarrow}{B}^{-1} \stackrel{\leftrightarrow}{C}$.

\subsection{Magnetic-field-dependent waves}

Now we turn to the magnetoplasma waves governed by eqs. (4). Invoking the
Bloch theorem for the field components ($E_x$ and $E_y$), making Fourier
expansion of the position-dependent material parameters ($\epsilon_{xx}$
and $\epsilon_{xy}$), and normalising the quantities, just as before, leads
us to write eqs. (4) in the form
\begin{myeqnarray}
\sum_{\vec{g}'} \left [ A(\vec{g}-\vec{g}') - (\vec{k}+\vec{g}')_y
(\vec{k}+\vec{g}')_y \delta_{\vec{g}'-\vec{g}} \right ] E_x(\vec{g}')\nonumber\\
+ \sum_{\vec{g}'}\left [ B(\vec{g}-\vec{g}') + (\vec{k}+\vec{g}')_x
(\vec{k}+\vec{g}')_y \delta_{\vec{g}'-\vec{g}} \right ] E_y(\vec{g}')=0\\
\sum_{\vec{g}'} \left [ B(\vec{g}-\vec{g}') - (\vec{k}+\vec{g}')_x
(\vec{k}+\vec{g}')_y \delta_{\vec{g}'-\vec{g}} \right ] E_x(\vec{g}')\nonumber\\
- \sum_{\vec{g}'}\left [ A(\vec{g}-\vec{g}') - (\vec{k}+\vec{g}')_x
(\vec{k}+\vec{g}')_y \delta_{\vec{g}'-\vec{g}} \right ] E_y(\vec{g}')=0
\end{myeqnarray}
where functional $A(\vec{g})$ and $B(\vec{g})$ are defined as
\begin{equation}
A(\vec{g})= \left \{
\begin{array}{lll}
A_i f + A_b (1-f) \equiv \overline {A}\ \ , \ \mbox{for} \ \vec{g} = 0 \\
(A_i-A_b)F(\vec{g}) \equiv \Delta A F(\vec{g})\ \ , \ \mbox{for} \ \vec{g} \ne 0
\end{array}
\right .
\end{equation}
and
\begin{equation}
B(\vec{g})= \left \{
\begin{array}{lll}
B_i f + B_b (1-f) \equiv \overline {B}\ \ , \ \mbox{for} \ \vec{g} = 0 \\
(B_i-B_b)F(\vec{g}) \equiv \Delta B F(\vec{g})\ \ , \ \mbox{for} \ \vec{g} \ne 0
\end{array}
\right .
\end{equation}
where $\vec{k}$ and $\vec{g}$ are just as defined before and
\begin{myeqnarray}
A_j = (\frac {a}{2\pi})^2 q^2_0 \epsilon^j_{xx}=\epsilon^j_L \Omega^2
\left ( 1-\frac{\Omega^2_{pj}}{\Omega^2-\Omega^2_{cj}} \right ) \\
B_j = (\frac {a}{2\pi})^2 q^2_0 \epsilon^j_{xy}= i\epsilon^j_L
\frac {\Omega \Omega_{cj}\Omega^2_{pj}}{(\Omega^2-\Omega^2_{cj})}
\end{myeqnarray}
with sub/superscript $j\equiv i,b$. Here $\Omega_{cj}=(\omega_{cj} a/2\pi c)$
is the normalized electron cyclotron frequency (see Appendix).

Let us now rewrite eqs. (18) symbolically as follows:
\begin{myeqnarray}
\stackrel{\leftrightarrow}{A_1}E_x(\vec{g}') +
\stackrel{\leftrightarrow}{A_2}E_y(\vec{g}')=0\\
\stackrel{\leftrightarrow}{A_3}E_x(\vec{g}') +
\stackrel{\leftrightarrow}{A_4}E_y(\vec{g}')=0
\end{myeqnarray}
where the matrices $\stackrel{\leftrightarrow}{A_j}; j \equiv 1, 2, 3, 4$;
are defined such that the matrix elements thereof are given by
\begin{myeqnarray}
A_{1(\vec{g},\vec{g}')}=A(\vec{g}-\vec{g}')-(\vec{k}+\vec{g}')_y
(\vec{k}-\vec{g}')_y \delta_{\vec{g}'-\vec{g}}\\
A_{2(\vec{g},\vec{g}')}=B(\vec{g}-\vec{g}')+(\vec{k}+\vec{g}')_x
(\vec{k}-\vec{g}')_y \delta_{\vec{g}'-\vec{g}}\\
A_{3(\vec{g},\vec{g}')}=-B(\vec{g}-\vec{g}')+(\vec{k}+\vec{g}')_x
(\vec{k}-\vec{g}')_y \delta_{\vec{g}'-\vec{g}}\\
A_{4(\vec{g},\vec{g}')}=A(\vec{g}-\vec{g}')-(\vec{k}+\vec{g}')_x
(\vec{k}-\vec{g}')_x \delta_{\vec{g}'-\vec{g}}
\end{myeqnarray}
Eliminating $E_y(\vec{g}')$ from eqs. (22) leaves us with
\begin{equation}
\left ( \stackrel{\leftrightarrow}{A_2}^{-1}\stackrel{\leftrightarrow}{A_1}-
\stackrel{\leftrightarrow}{A_4}^{-1}\stackrel{\leftrightarrow}{A_3}\right )
E_x(\vec{g}')=0
\end{equation}
Nontrivial solutions of this equation are given by
\begin{equation}
\left |
\stackrel{\leftrightarrow}{A_2}^{-1}\stackrel{\leftrightarrow}{A_1}-
\stackrel{\leftrightarrow}{A_4}^{-1}\stackrel{\leftrightarrow}{A_3}
\right | =0
\end{equation}
What is noteworthy is the fact that in this magnetic-field-dependent part of
the problem, the secular equation cannot be cast in the form of a standard
eigenvalue problem. That is because the eigenvalues are involved in both the
diagonal and nondiagonal elements of the resultant matrix. So the only way
down to the determination of the eigenvalues is to treat eq. (25) as a
transcendental function and find its zeros for a given $\vec{k_\parallel}$, by
inserting an additional loop over $\Omega$. Naturally, the time consumption
during the computation of reliable band structures will be enormous.

It is interesting to note that one can, without much efforts, prove
analytically that, in the limit of vanishing applied magnetic field
($\Rightarrow$ functional $B(\vec{g}) \rightarrow$ 0), the secular equation
(25) reduces to
\begin{equation}
\left | \stackrel{\leftrightarrow}{A}- \Omega^2 \stackrel{\leftrightarrow}{I}
\right | =0
\end{equation}
where $\stackrel{\leftrightarrow}{I}$ is a unit matrix. Eq. (26) is exactly
identical to the secular equation obtained from eq. (14). This emboldens our
confidence in the adequacy of the formalism presented in this work.

The last step in presenting the numerical examples is to specify the symmetry
of the lattices. The specific geometrical arrangements considered in this work
are the square and hexagonal lattices. Their reciprocal lattice vectors are
given by
\begin{equation}
\vec{G}=\frac {2\pi}{a}(n_x\hat{x}+n_y\hat{y})
\end{equation}
for the square lattice and
\begin{equation}
\vec{G}=\frac{2\pi}{a}\left [n_x\hat{x}+((-n_x+2n_y)/\surd{3})\hat{y}\right ]
\end{equation}
for the hexagonal lattice. In this work the integers $n_x$ and $n_y$ were
limited in the range of $-10 \leq (n_x,n_y)\leq +10$; which implies to 441
plane waves considered. This resulted in a reliably very good convergence; at
least up to the first fiftieth bands. In this paper we will confine ourselves
to the waves which turn out to be independent of the applied magnetic field
and hence are governed  by eq. (14). The analytical and numerical details of
the magnetic-field dependent waves are deferred to a forthcoming publication.

\section{NUMERICAL EXAMPLES}

For the purpose of numerical computation, we specify the semiconductor-
dielectric composite to be made up of doped GaAs--intrinsic GaAs or doped
GaAs--vacuum. This implies that $\epsilon_L$ = 12.8 (1.0) for doped or
undoped GaAs (vacuum). As stated above, we confinr ourselves to the waves
which are independent of the applied magnetic field. As such the only other
parameter needed to accomplish the computation is the dimensionless plasma
frequency $\Omega_p$; two specific values used in this work are $\Omega_p$=
0.5 and 0.75. We will discuss a variety of illustrative numerical results.
This includes exchanging materials in the cylindrical inclusions and the
background medium; in addition to the two values of $\Omega_p$. The whole
range of filling fraction is explored in order to discuss the regime of at
least first ten bands of the respective composites in both square and
hexagonal lattices (see Fig. 1).

\subsection{Square lattice}

Figure 2 illustrates the band structure (BS) and density of states (DOS) for
a square lattice of doped GaAs cylinders in vacuum for a filling fraction
$f=10\%$. The plots are rendered in terms of the dimensionless frequency
$\Omega$ versus Bloch vector $\vec{k}$. The left part of the triptych
represents the BS in three principal symmetry directions, letting $\vec{k}$
scan only the periphery of the irreducible part of the first Brillouin zone
(see the inset on the right). The middle part is the result of an extensive
scanning of $\mid \vec{k} \mid $ in the irreducible triangle $\Gamma$XM
of the Brillouin zone --- the interior of this zone and its surface , as
well as the principal directions shown in the left part of this figure. Each
curve here corresponds to some direction of $\vec{k}$. The DOS in the right
part of the triptych has been calculated on the basis of the scanning in the
middle part, which corresponds to 1326 $\vec{k}$ points. The three parts of
the triptych together demonstrate that there is, indeed, a genuine full gap
in the frequency range defined by $0\leq \Omega \leq 0.320$ and we consider
such calculations as essential.

Extensive computation of the BS for the same composite as considered in Fig.
2 reveals that one can also achieve some higher energy gaps within the first
ten bands  for certain values of the filling fraction. For instance, for
$f=30\%$, there are three full gaps within the first ten bands --- the first
one is analogous to the one shown in Fig. 2, the second one exists between the
third and fourth bands, and the third one opens between the ninth and tenth
bands. For $f=40\%$, the third gap exists between the sixth and seventh bands,
whereas the second gap, although differs in magnitude and frequency range, still
persists between the third and fourth bands. For $f=50\%$, there are only two
gaps --- the first one exists below the first band and the second one opens
between the sixth and seventh. We find that, whether or not the higher energy
gaps exist, the lowest gap below a threshold frequency $\Omega_t$
(which varies with $f$) defined as the minimum of the first band at the
$\Gamma$ point, always persists. It has been noted that, for $f\geq 60\%$,
only the lowest gap exists. This is true even for the close-packing value
($f=0.7853$) where the cylinders just touch each other.

The dependence of the gap-widths of the lowest (which is also the widest) gap
on the filling fraction for diversely  designed composites in square lattice
is summarized in Fig. 3. The curves designated as ASj, SAj, ISj, and SIj refer
, respectively, to the composites made up of cylindrical holes in doped GaAs
background, cylindrical doped GaAs in vacuum, cylindrical intrinsic GaAs in
doped GaAs background, and cylindrical doped GaAs in intrinsic GaAs background;
$j=1(2)$ correspond to $\Omega_p=0.5(0.75)$. The important points one can
notice at glance are the following. The composites made up of doped
semiconductor inclusions in the dielectric background give rise to the lowest
gap (below $\Omega_t$) whose width increases with increasing filling fraction
(see, for example, the curves SAj and SIj; $j = 1,2$). On the other hand, the
composites made up of dielectric cylinders in the doped semiconductor background
give rise to the lowest gap which decreases as the filling fraction increases
(see, for example, the curves ASj and ISj). Comparing the curves ASj with ISj
(or SAj with SIj) leads us to infer that the larger the dielectric mismatch
(between the inclusions and the background), the wider is the gap-width.
Similarly, comparing the respective curves for j=1 with those for j=2 reveals
that the gap-width becomes larger as the plasma frequency increases. It is also
noteworthy that the maximum gap-width (of the lowest gap) one can achieve is,
empirically, defined by $ 0\leq \Omega_t \leq \Omega_p$. This implies that the
largest gap to midgap ratio (GMR) one can achieve is exactly two. This abides
by the fact that in order to obtain GMR $\geq 1$, $\Omega_2$ should be
$\geq$ $3\Omega_1$; $\Omega_2$ ($\Omega_1$) being the frequency of the top
(bottom) of the existing gap.

\subsection{Hexagonal lattice}

Figure 4 ilustrates the band structure and DOS for a hexagonal lattice of doped
GaAs cylinders in vacuum for a filling fraction $f=10\%$. The plots are rendered
in terms of the dimensionless frerquency $\Omega$ versus Bloch vector $\vec{k}$.
The left part of the triptych represents the BS in three principal symmetry
directions, letting $\vec{k}$ scan only the periphery of the irreducible part of
the first Brillouin zone (see the inset on the right). The middle part is the
result of an extensive scanning of $\mid \vec{k} \mid $ in the irreducible
triangle $\Gamma$JX of the Brillouin zone --- the interior of this zone and its
surface , as well as the principal directions shown in the left part of this
figure. The DOS in the right part of the triptych has been calculated on the
basis of the scanning in the middle part, which corresponds to 1275 $\vec{k}$
points. The three parts of the triptych together demonstrate that there is a
genuine full gap in the frequency range defined by $0\leq \Omega \leq 0.329$. In
addition, there is another band-gap opening up between the first and second bands
--- defined by the maximum (minimum) of the first (second) band at J (X) point.

Extensive computation of the BS for the same composite as represented by Fig. 4
reveals that there are some other higher energy gaps opening up within the first
ten bands for certain values of $f$. For instance, for $f=20\%$, there are three
more full gaps, apart from the lowest one that opens below $\Omega_t$ --- the
first one exists between the first two bands, the second one opens between the
third and fourth bands, and the third one exists between the eighth and ninth
bands. For $30\% \leq f\leq 60\%$, we obtain two more full gaps, apart from the
lowest one, with varying frequency range and magnitudes. At $f=70\%$, we left
with only two full gaps --- the first one is the same that exists below the
minimum (at $\Gamma$ point) of the lowest band and the second one opens between
the third and fourth bands. For $f\geq 70\%$, there are no more full gaps except
the lowest one. The semiconductor-dielectric composites in the hexagonal lattice,
just as in the square lattice, are always seen to give rise to the lowest gap
persisting below a threshold frequency $\Omega_t$.

The dependence of the lowest gap-width on the filling fraction for diversely
designed composites in the hexagonal lattice is depicted in Fig. 5. Comparing
Fig. 5 with Fig. 3 clearly reveals that the gap-widths (of the lowest gap) for
the respective composites in the hexagonal lattice, over almost the whole
common range of filling fractions, are larger than those in the square lattice.
The rest of the discussion related to Fig. 3 is still seen to be valid. It should
be pointed out that this trend of achieving larger gaps in the hexagonal lattice
as compared to the square lattice is the same as has been noted earlier in the
cases of photonic as well as phononic crystals$^4$. This seems to be physically
quite a reasonable result, because the constant energy surfaces in the hexagonal
lattice are closer to the circular shape than those of a square lattice.

\section{CONCLUDING REMARKS}

In summary, we have demonstrated that the simple 2D periodic binary semiconductor-
dielectric composites can give rise to full plasmonic band-gaps. Although we consider
an applied magnetic field oriented along the axis of the cylindrical inclusions, two
independent polarizations of the waves have been obtained --- $E_z \parallel\hat{z}$
axis and $ H_z \parallel\hat{z}$ axis, just as in the absence of an applied magnetic
field. Only the waves in the latter polarization are magnetic-field dependent. The
numerical results presented in this work correspond to the former polarization. As
such the term "full" should be reserved with respect to this polarization. In principle,
the band structures computed separately for the two polarizations should be superimposed
--- one over the other --- to check and claim whether or not the band-gaps are
independent of the polarization. This is the task aimed at in a future  publication.

Our theoretical results on achieving full band-gaps must be accompanied by a word of
warning. Apart from the fact that both polarizations need to be considered, we have
here also assumed the z-component of the wave vector ($k_z$) to be zero. Considering
$k_z$ finite would give a better insight into the problem. We have also ignored the
damping effects and the interaction of the charge carriers with the optical phonons.
The latter could be taken into account by considering the frequency-dependence of
the background dielectric constants. Additional work addressing the effect stemming
from the cylinders of different cross-sections and orientation of the external magnetic
field would be worthwhile. Considering the cylindrical inclusions of finite lengths
should be of more practical interest.

Attractive possibilities exist for the experimental observation of such plasmonic
band-gaps, both with and without an applied magnetic field, in the semiconductor-
dielectric composites as investigated here. Resonant Raman scattering, as employed
to study plasmons in quantum wire superlattices, may be one option worth attempting.

\acknowledgements
The author would like to thank F. Perez-Rodriguez for useful discussions.
This work was partially supported by CONACYT Grant \# 2373--PE.

\section{APPENDIX}

The dielectric tensor components employed in this work are defined as follows:
\begin{myeqnarray}
\epsilon_{xx}=\epsilon_{yy}=\epsilon_L\left [ 1-\frac{\omega^2_p(\omega+i\nu)}
{\omega \left [(\omega+i\nu)^2-\omega^2_c\right ]} \right ]\\
\epsilon_{yx}=-\epsilon_{xy}=-i\epsilon_L \frac{\omega^2_p \omega_c}
{\omega \left [ (\omega+i\nu)^2-\omega^2_c\right ]}\\
\epsilon_{zz}=\epsilon_L \left [ 1-\frac {\omega^2_p}
{\omega (\omega+i\nu)}\right ]
\end{myeqnarray}
where $\epsilon_L$ is the background dielectric constant, $\nu$ is the free carrier
collision frequency, $\omega_p$ is the screened plasma frequency, and $\omega_c$
is the electron cyclotron frequency.

If we also consider the effect of phonons, which, in a way, incorporates the
coupling of the plasmons (or magnetoplasmons) to the optical phonons, then
the background dielectric constant $\epsilon_L$ has to be replaced by its
frequency dependent expression:
\begin{equation}
\epsilon_L(\omega)=\epsilon_{\infty}\left [\frac {\omega^2_{LO}-\omega^2-i\Gamma\omega}
{\omega^2_{TO}-\omega^2-i\Gamma\omega}\right ]
\end{equation}
where $\epsilon_{\infty}$ is the high-frequency dielectric constant, $\Gamma$ is
the optical-phonon damping frequency, and $\omega_{LO}$ and $\omega_{TO}$  are,
respectively, the longitudinal and transverse optical phonon frequencies at
the zone center of the Brillouin zone.

\newpage
\noindent{\Large\bf REFERENCES}

\bigskip
\noindent\begin{tabular}{rp{16.5cm}}
$^1$ &  See, e.g., {\it Proceedings of the $11^{th}$ International Conference
        on Electronic properties of Two-Dimensional Systems (EP2DS)},
        Surf. Sci. {\bf 361/362} (1996).\\
$^2$ &  See, e.g., P. Rai-Chaudhry, Editor, {\it The Handbook of
        Microlithography, Micromachining, and Microfabrication}
        (SPIE, 1996).\\
$^3$ &  E. Yablonovitch, Phys. Rev. Lett. {\bf 58}, 2059 (1987);
        S. John, {\it ibid}. {\bf 58}, 2486 (1987).\\
$^4$ &  For a recent extensive review, see M.S. Kushwaha, Int. J.
	       Mod. Phys. B{\bf 10}, 977 (1996).\\
$^5$ &  M.S. Kushwaha, P. Halevi, L. Dobrzynski, and B> Djafari-Rouhani, Phys. Rev. Lett. {\bf 71}, 2022(1993); M.S. Kushwaha, Appl. Phys. Lett. {\bf 70}, 3218 (1997); and
        references therein.\\
$^6$ &  J.O. Vasseur, L. Dobrzynski, and B. Djafari-Rouhani, Phys. Rev.
        B {\bf 54}, 1043 (1996).\\
$^7$ &  S. John, Phys. Rev. Lett.{\bf 53},2169 (1984).\\
$^8$ &  A. Kobayashi and R.E. Prange, Phys. Rev. Lett.
        {\bf 56}, 1280 (1986).\\
$^9$ &  A.R. McGurn and A.A. Maradudin, Phys. Rev. B{\bf 48}, 17576
           (1993); see also V. Kuzmiak and A.A. Maradudin, Phys. Rev. B {\bf 58}, 7427(1997).\\
$^{10}$ &  R.F. Wallis and J.J.Quinn, Phys. Rev. B {\bf 38}, 4205
           (1988).\\
$^{11}$ & M.S. Kushwaha, Phys. Rev. B{\bf 42}, 5602 (1990). \\
$^{12}$ & B.L. Johnson and R.E. Camley, Phys. Rev. B {\bf 43}, 6554
          (1991).\\
$^{13}$ & R. Strenz et al. Phys. Rev. Lett. {\bf 73}, 3022 (1994).\\
$^{14}$ & M.S. Kushwaha and P. Zielinski, Europhys. Lett.
          (to be published).\\
\end{tabular}

\begin{figurecaptions}

\item   Schematics of a few units cells of square (a) and hexagonal (b)
        lattices investigated in this work. The cylindrical inclusions are
        oriented along the $\hat{z}$ axis. The propagation is confined to the
        $\hat{x}-\hat{y}$ plane. An external magnetic field field is orinted
        parallel to the cylinders.
       
\item   Band structure (BS) and density of states (DOS) of a square lattice of
        cylindrical (GaAs) semiconductor inclusions of circular cross-section
        in a vacuum background. The filling fraction is $f=10\%$. The triptych
        is comprised of three parts. In the left panel, we plot the BS in three
        principal symmetry directions, letting $\vec{k}$ scan only the
        periphery of the irreducible part of the first Brillouin zine (see
        inset on right). The middle panel demonstrates a novel way of plotting
        the eigenvalues as a function of $\mid \vec{k} \mid $; i.e., the
        distance of a point in the irreducible part of the Brillouin zone from
        the $\Gamma$ point. The right panel illustrates the DOS. We call
        attention to the full gap below a threshold frequency $\Omega_t$=0.320
        and extending down to zero.

\item   Dependence of the gap-widths (of the lowest gap) on the filling fraction
        for diversely designed composites in a square lattice. The curves
        designated as ASj, SAj, ISj, and SIj refer, respectively, to the
        composites made up of cylindrical holes in doped GaAs background,
        cylindrical doped GaAs in vacuum, cylindrical intrinsic GaAs in doped
        GaAs background, and cylindrical doped GaAs in intrinsic GaAs background.
        The horizontal dashed lines stand for the plasma frequencies $\Omega_{p1}
        =0.50$ and $\Omega_{p2}=0.75$.

\item   The same as in Fig. 2, but for the hexagonal lattice. The threshold
        frequency for the lowest gap that extends down to zero is
        $\Omega_t=0.329$. Notice a small second full gap (hatched region)
        existing between the first two bands.

\item   The same as in Fig. 3, but for the hexagonal lattice.
  
\end{figurecaptions}
\end{document}